\documentclass[12pt,halfline,a4paper]{ouparticle}
\usepackage{graphicx, amssymb, amsmath} 
\usepackage{color}
\usepackage{caption}

\newcommand {\de}{\delta}

\newcommand {\al}{\alpha}

\newcommand {\ka}{\kappa}
\newcommand {\fr}{\frac}

\newcommand {\beg}{\begin{equation}}
\newcommand {\en}{\end{equation}}
\newcommand {\bega}{\begin{eqnarray}}
\newcommand {\ena}{\end{eqnarray}}

\usepackage{graphicx}
\usepackage{lmodern}
\usepackage[T1]{fontenc}
\usepackage[english,activeacute]{babel}
\begin{document}

\title{A hydrodynamic analytical model of fish tilt angle: Implications for the modelling of the acoustic target strength}

\author{\name{F. Agust\'{i}n Membiela}\email{agustinmembiela@gmail.com} 
\address{IFIMAR (CONICET and Universidad Nacional de Mar del Plata),
De\'an Funes 3350, B7602AYL Mar del Plata, Argentina.}
\and
\name {Mat\'{i}as G. dell'Erba}
\email{mdellerba@ifimar-conicet.gob.ar}
\address{IFIMAR (CONICET and Universidad Nacional de Mar del Plata),
De\'an Funes 3350, B7602AYL Mar del Plata, Argentina.}}

\abstract{
We implement a simple hydrodynamical model to study behavioural swimming tilt angle of open swimmbladder fish. For this purpose we study the stability of forces acting on a fish swimming horizontally with constant velocity. Additionally, the open swimbladder compression with the depth is modelled by Boyle's law. With these, our model gives an analytical solution relating the depth with the body tilt angle and the velocity. An interesting result for steady horizontal swimming is that the body tilt decreases with velocity almost like $v^{-1}$. Moreover, we give an expression for the maximum tilt angle. Then, by introducing the assumption of constant swimming power we relate the swimming velocity with the tilting. Furthermore, we show that the hydrodynamical influence of a temperature gradient produced by a thermocline seems to be negligible for the fish tilting. These results are considerably helpful for more realistic modelling of the \emph{acoustic target strength of fish}. Finally, we tested our results by comparing the hydrodynamics solutions with others obtained from acoustic observations and simulations of target strength for Argentine anchovy.}


\maketitle

\section{Introduction}


Teleost fish use their swimbladders to gain buoyancy. In this regards they could be divided into physoclists, with a closed swimbladder and physostomes with an open swimbladder (Morrisey et al., 2016; Gilles, 1991). The physoclist fish have a gas secretion system bind to their bloodstream through which they can adjust the swimbladder volume. This allows them to keep close to neutral buoyancy for different depths. Besides, the physostomous fish, that posses a more primitive morphology, are incapable of refilling their swimbladders to avoid an intense negative buoyancy when going to deeper depths. In this case, the swimbladder seems only to be refilled by gulping atmospheric air from the surface (Blaxter and Hunter, 1982; Blaxter and Batty, 1984; Brawn, 1962). 

It is well known in ecological acoustics that the primary reflection of acoustic energy happens at the swimbladder (Foote, 1980). Besides, fish with swimbladders comparable or longer than the wavelength are acoustically directive. Thus, small changes in the tilt of the fish can affect significantly the measured target strength (TS). In Hazen and Horne (2003) et al. it was shown that the influence of tilt on TS is greater than the fish length change. Indeed, an incorrect estimation of behavioural tilt angle distribution may spoil the determination of biomass from echoenergy integration. The variation of TS as a function of the tilt angle has been extensively studied (Foote, 1980; Huse and Ona, 1996; Foote, 1980b; Nakken and Olsen, 1977, McQuinn and Winger, 2003; Prario et al., 2015, Kang et al., 2005). On the other hand, it was suggested that negatively buoyant fishes may adopt a positive body tilt during steady horizontal swimming as a behavioural mechanism to avoid sinking (Huse and Ona, 1996; He and Wardlet, 1986; Wilga and Lauder, 2000). Moreover, in Strand et al. (2005), they discussed the energy saving obtained through fish swimming tilt angle for physoclist.

The theoretical modelling of fish target strength involves the previous knowledge of fish behavior, i.e., the tilt angle distribution of fish. This is necessary to properly average the target strength functions over a certain tilt angle distribution. However, this knowledge is rarely fulfilled since direct observations of the fish swimming tilt angle are scarce. This is specially so when the aim of the TS models is to provide insight on the acoustical scattering of the organisms in the wild, which in turn constitutes a major concern for the acoustic assessment surveys. We aim to develop a simple model of the fish swimming to be used to estimate the "expected" tilt angle adopted by a particular fish species. We consider the influence of certain external factors influencing fish behavior and we concentrate on the particular case of physostomous fish, i.e. fish lacking a swimbladder, accomplishing vertical migrations in the water column and hence exposed to a significant variation of the hydrostatic pressure and the consequent effects on its buoyancy conditions. The specific anatomy of fish is considered in order to provide an approximate but more realistic source for estimating the average tilt angle of fish. In this way we expect to narrow the subjectivity usually associated to the selection of fish tilt angle pdf's for averaging TS functions.

On the other hand, several authors have studied fish swimming focusing in the hydrodynamics of undulatory propulsion generation (Sfakiotakis et al., 1999; Shadwick and Lauder, 2006). Nowadays, the preferred simple analytical model of fish swimming kinematics is Sir J. Lighthill's elongated body theory (Lighthill, 1960; Lighthill, 1971), which is based on adding up the forces due to the lateral acceleration and deceleration of the body as it undulates from side to side.

We are interested in developing a minimal model by studying simple hydrodynamics of the body and the pectoral fins, independently of the propulsion movements and thrust generation, so as to obtain general results within a minimal set of parameters. Some benefits of our approach are that we will obtain analytical expressions, that may offer a deeper insight, and that are also applicable to a variety of species. 

Though the model we shall develop is far from giving a thoroughly cover of fish swimming, it is aimed as a first minimal model to analyse tilting behaviour for physostomous fish and may be easily extended to physoclists and negative buoyant fish without swimbladder. Specifically, we have chosen a simplified swimming behaviour by assuming a steady horizontal trajectory, in contrast with other swimming strategies, like \textit{glide and rise}, which may be adopted for deep depths (Huse and Ona, 1996; Tanaka et al., 2001).

By means of physiological and hydrodynamical arguments, we will try to shed some insight on the relation  between the tilt angle, the swimming depth and the swim velocity of physostomous fish that suffer loss of buoyancy because of swimbladder compression with depth. This will help on the construction of simulations of individual TS and volume scattering of fish schools.

We ordered the paper in the following manner. In section \ref{sec:2}, we introduce the hydrodynamical model from which we obtain a simplified equation relating the tilting angle with the depth and the swimming velocity. In turn, we add a hypothesis of constant swimming power which yields a tilting dependence on the velocity. Additionally, we consider the effects of the presence of a thermocline in the model. In section \ref{acoustic}, we test our model against acoustic data and simulations of TS of Argentine anchovy. Finally, in section \ref{conclu} we discuss the results and give our conclusions.

\section{The model}\label{sec:2}

We shall study the hydrodynamic vertical stability of a fish swimming at stationary velocity and depth, for which we consider the different forces acting on the fish components (fins, body, etc.). In this approach we will search for general results independent of the constraints introduced by means of the bioenergetic costs and the efficiency of buoyancy regulation during swimming. This last common bioenergetics framework has been modelled by Strand et al. (2005) for physoclists, by including the trade-offs between the swimbladder regulation and the hydrodynamics forces. 

In addition, fish are self-propelled through undulatory movements, thus it is hard to separate thrust generation of their inherited drag forces (Schultz and Webb, 2002). However, if we assume that these movements are confined to a reduced portion of the posterior fish body and/or the caudal fin, then we can approximate the fish body as a rigid body. Indeed, in such cases we can consider that the fish body behaves hydrodynamically similar to a torpedo-shaped body travelling with a definite tilt angle (Jorgensen, 1973; Evans, 2003). In this way, we shall consider the simplification that the interactions between the fluid and the fish body are independent of the thrust generating undulatory movements.

For the experimental validation of this model, its variables (such as velocity, depth and tilt) represent averages from the experimental data. However, in a first approach this validation may also be performed by numerical simulations that model the complete hydrodynamic picture. For example, this can be done for thunniform and carangiform swimmers (Sfakiotakis et al., 1999) by introducing the hydrostatics of the open swimbladder. We expect that corrections may indeed appear to our framework when including the undulatory phenomenology, the thrust generation and the swimming bio-energetics.

\subsection{Forces during stationary swimming}

\begin{center} 
\begin{figure}
\includegraphics[width=10cm]{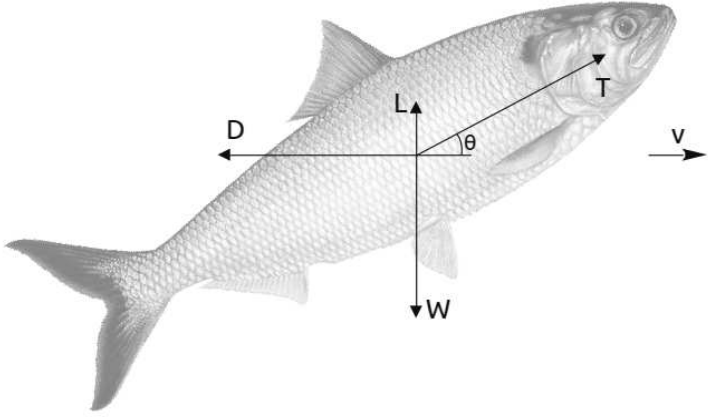}
\caption{Forces acting on a fish during horizontal swim with velocity $v$: $\mathbf{T}$, the thrust force in the direction $\theta$; $\mathbf{W}$, the apparent weight; $\mathbf{L}$, the total lift force, and $\mathbf{D}$ the total drag force.}
\label{pez}
\end{figure}
\end{center}

In Figure \ref{pez} we summarized the forces acting on the fish while swimming horizontally with velocity $v$ and tilt $\theta$. The thrust force $\mathbf{T}$ is the consequence of the propulsion obtained from the specific fish swimming mode. In the carangiform and the thunniform swimming modes the thrust is obtained from the body and caudal fin, thrown into a wave with as much as up to one half-wavelength within the length of body (Webb, 1975). From the torpedo-body-shape assumption we expect that our study may only be limited to these undulatory modes.  In turn, swimming with stationary velocity and depth, determines a particular swimming strategy given by stationaries tilt and thrust. Other swimming strategies can be studied where the tilt and thrust are not constant (Videler, 1993; Taylor et al., 2010 ; Sfakiotakis et al., 1999).

The apparent weight $\textbf{W}$ is a force that always appears in submerged bodies, which is the resultant between the Archimedes force and the weight of the fish. This force changes with depth as the swimbladder compresses. If the fish has negative apparent weight at the surface (with $\textbf{W}$ pointing upwards), it means that floats on water, then in order to gain depth it will need to nose downwards with a negative tilt angle. Eventually, at a certain depth the average fish density equals the water density reaching the neutral buoyant depth. However, this is an unstable point since below of this depth $\textbf{W}$ changes sign, pointing downwards and the fish starts to sink. Then, for the fish to swim at a constant depth, this positive apparent weight should be compensated by some other vertical forces. Essentially there will be two kinds, lift forces and the vertical component of the thrust force for a swimming tilt angle. Hydrodynamical lift $\textbf{L}$ while swimming may be obtained mainly from pectoral fins acting as hydrofoils and from body lift. Additionally, hovering behaviour may be used but in this case the fish is not moving and cannot be combined with the other forces. Finally, the drag forces $\textbf{D}$, that always oppose to the flow direction, manifest as pressure drag and friction drag.  

There is an extensive bibliography on the study of the hydrodynamics (and aerodynamics) acting either on biological creatures or engineering systems (Webb, 1975; Vogel, 1996; Webb, 1988; Norberg, 1990; FAA, 2012; McNeill, 1982).

The steady motion assumption implies that all forces balance, thus there is no acceleration. From Newton laws we obtain,
  \bega
    L-W+T \sin\theta&=&0\\
    T\cos\theta-D&=&0,\label{Newton2}
 \ena
where $\theta$ is the angle between oblique thrust $\textbf{T}$ and horizontal velocity $\textbf{v}$. We can combine both equations to get,
 \beg\label{Newton}
    W=L-D\tan\theta.
 \en
In this paper we shall keep as simple as possible and we shall ignore the moments stability analysis. However, a more detailed study on particular fish species should incorporate this issue.

\subsection{Swimmbladder and body compression}
Some of the strategies of Nature to achieve buoyancy are the lipid (or oil) reservoirs in the fish flesh, which help to reduce the averaged fish density $\rho_b$ (Shadwick and Lauder, 2006; Alexander, 1990). In turn, other fish posses the gas-filled swimbladder organ that aims to the same purpose.
We shall consider that the open swimbladder follows Boyle's Law contraction rate (Gorska and Ona, 2003; Nero and Thompson, 2004) which means that
 \beg
     V_{sb}(z)=V_{sb0}\left(1+\frac{\rho g}{P_0} z\right)^{-1}
 \en
where $z$ is the depth below water surface, $V_{sb}(z)$ is the swimbladder volume at a depth $z$ (with $V_{sb}(0)=V_{sb0}$), $\rho$ is the water density, $g$ is the gravity acceleration and $P_0$ is the atmospheric pressure at sea level. Moreover, we assume that the mean flesh density is practically constant in spite of the swimbladder contraction with the depth. Thus, the loss of volume by the swimbladder is followed by an equal loss of external volume in the fish body $V_{sb}(z)-V_{sb0}=V_b(z)-V_{b0}$ ,where $V_b(z)$ is the volume of the entire fish at depth $z$. Using these considerations, we obtain an expression for the apparent weight depth dependence,
 \beg\label{weight}
    W(z)=\rho g V_{sb0}\left(1+\fr{z}{z_\ast}\right)^{-1}-(\rho_b-\rho) g (V_{b0}-V_{sb0}),
 \en
where  $\rho_b$ is the mean fish density and $\rho g z_\ast=P_0$, then $z_\ast\simeq10m$. Thus, the apparent weight increases hyperbolically from 
 \beg
	W_0=(\rho_b-\rho)g V_{b0}-\rho_b g V_{sb0},
  \en
at sea level ($z=0$), to deep values where the swimbladder is totally collapsed
 \beg
 	W_\infty=(\rho_b-\rho)g (V_{b0}-V_{sb0}),
 \en 
for $z\gg z_\ast$. The difference between these values is $W_\infty-W_0=\rho g V_{sb0}$. We can further define the \textit{neutral buoyancy depth} $z_N$ obtained from the condition $W(z_N)=0$. It is easy to see that,
 \beg
 	z_N=-z_\ast\fr{W_0}{W_\infty}.
 \en
As a consequence, when the fish flesh is denser than the sea water, $W_\infty>0$ and then in order to obtain positive buoyancy, $W_0<0$.

\subsection{Tilting depth dependence}\label{Tilting}

By combining equations (\ref{Newton}) with (\ref{weight}) we obtain
  
  
  \begin{equation}\label{zeta}
    z(\theta ,v)=z_\ast\left(\frac{1}{F(\theta,v) +f_0}-1\right),
  \end{equation}
where we have defined,
 \bega
 F(\theta,v)&=&-\frac{D(\theta ,v) \tan\theta+  L (\theta,v)}{\rho g V_{sb0} } \simeq v^2 f_1(\theta),\label{v2f1} \\
 f_0&=& \left(\frac{\rho_b }{\rho}-1\right) \left(\frac{V_{b0}}{V_{sb0}}-1\right).\label{f1f2}
 \ena

Equations (\ref{zeta}-\ref{f1f2}) are the main results of the paper, they give the depth at which the forces on the fish balance, for a steady horizontal swimming with velocity $v$ and tilting $\theta$. In Appendix A we present some expressions for the estimation of $\mathbf{D}$ and $\mathbf{L}$ forces and we discuss the validity of the factorization $v^2 f_1(\theta)$ in Equation (\ref{v2f1}). It is not possible, though, to obtain an exact analytical expression $\theta=\theta(z,v)$ from Equation (\ref{zeta}). We have checked that $f_1(\theta)$ behaves practically like a quadratic function in the range of studied parameters. Therefore, by keeping to order $\theta^2$ in the Taylor expansion of $f_1(\theta)$ we obtain

\begin{equation}\label{f1}
f_1(\theta) \simeq -\frac{1}{2 g V_{sb0}}\left[ 2(S_f k + A_s)\theta + \bar{C}_{\perp} A_p \theta^2\right], 
 \end{equation}
where the parameters in the expansion are detailed in Appendix A. 

\begin{figure}
\includegraphics[width=10cm]{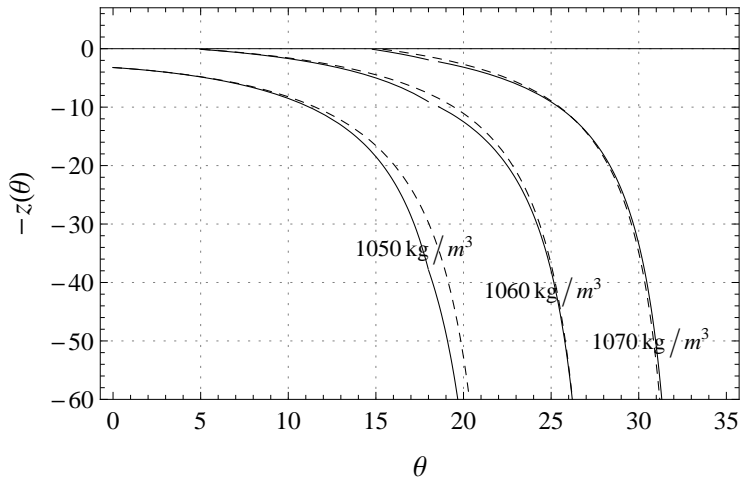}
\caption{$-z[m]\hspace{0.1cm} \rm{vs.} \hspace{0.1cm}\theta[deg.]$ from (\ref{zeta}) (filled) and with the approximation (\ref{f1}) (dashed) for velocity $v=1.0 \ell/s$ and densities $\rho_b=1050 kg/m^3$, $\rho_b=1060 kg/m^3$ and $\rho_b=1070 kg/m^3$.}\label{apden}
\end{figure}

In Figure (\ref{apvel}) we compare the functions $z(\theta)$ from the exact expression Equation (\ref{zeta}) (solid line) with the approximation from Equation (\ref{f1}) (dashed line) for swimming velocities $v[\ell/s]=0.8, 1.0, 1.5$ and fish flesh density $\rho_b[kg/m^3]=1060$. The parameters used from now on correspond to a representative specimen of Argentine anchovy and they are given in the table of Sec.\ref{param}. We remark that the approximation is good within a maximum error of about $1^\circ$. Furthermore, in Figure (\ref{apden}) we obtain similar results for $z(\theta)$ for different densities $\rho_b[kg/m^3]=1050, 1060, 1070$ and $v[\ell/s]= 1.0$. 

 \begin{figure}
\includegraphics[width=10cm]{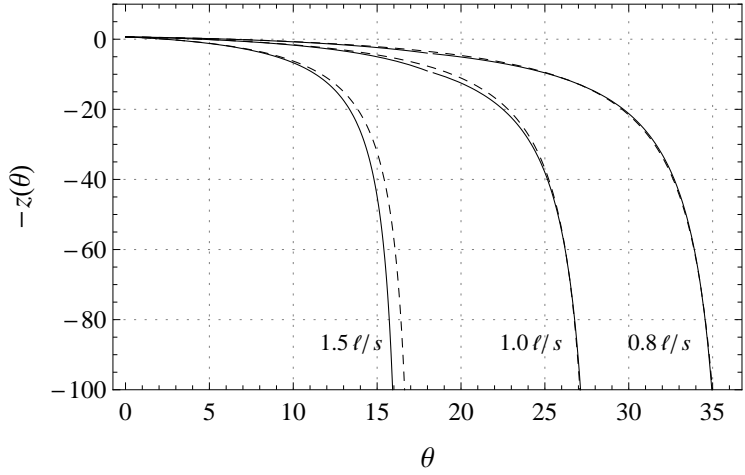}
\caption{$-z[m]\hspace{0.1cm} \rm{vs.} \hspace{0.1cm}\theta[deg.]$ from (\ref{zeta}) (filled) and with the approximation (\ref{f1}) (dashed) for body density $\rho_b=1060kg/m^3$ and velocities $v=0.8 \ell/s$, $v=1.0 \ell/s$ and $v=1.5 \ell/s$.}\label{apvel}
\end{figure}

Using Equations (\ref{zeta}-\ref{f1}) we can solve the tilting angle depth dependence  

 \begin{eqnarray}\label{theta_z}
\theta (z,v) &\simeq& \frac{S_f k + A_s}{A_p \bar{C}_{\perp} }\times \\
\nonumber&& \left\{ 1 + \sqrt{1+\frac{2 g V_{sb0} A_p\bar{C}_{\perp}}{(S_f k + A_s)^2}\left[\left(\frac{\rho_b }{\rho}-1\right) \left(\frac{V_{b0}}{V_{sb0}}-1\right)-\frac{z_\ast}{z_\ast+z}\right] \frac{1}{v^2}} \right\}.
 \end{eqnarray}
 
It is immediate to see that the tilting $\theta$ increases with the depth to a maximum asymptotic value $\theta_{\textrm{max}}$ for $z\gg z_\ast$. The value of $\theta_{\textrm{max}}$ is given by the last expression without the $z$ term. The tilting decreases at an approximate rate $\sim v^{-1}$ as could be seen in Equation (\ref{theta_z}). This happens because the main contribution comes from $\theta^2$ term in Equation (\ref{f1}) that belongs to the planform area of the fish. This result is compatible with experimental fits $\theta(v)\propto v^{-1.14}$ obtained for trout in Webb (1993). Thus, the faster the fish swims the less tilt it needs to achieve stability, because at faster movement the lift forces grow, allowing to reach the body stability with less tilt (Figure (\ref{apvel})) (He and Wardlet, 1986; Svendsen et al., 2005). Otherwise, the tilting will be enhanced for denser fishes, because an extra component of vertical thrust is needed to cancel the extra weight (Figure (\ref{apden})). In the last plot we can see how for more negative buoyant fish, the body tilt at the sea surface increases. In particular, we observe that for a flesh density $1050kg/m^3$ the buoyancy on surface is positive because the averaged density of the fish including the swimbladder is lower than the water density. In this case the fish should either swim with negative tilt or lose some air of the swimbladder. 

The expression Equation (\ref{theta_z}) simplifies considerably when the fish pectoral fins do not generate lift, either because of behavioural reasons (retracted fins during swimming) or morphological limitations $k S_f\ll A_p$  and when the tail section is small $A_p\gg A_s$, . Thus, we can only keep with the last term in the root square and the tilting decays like $v^{-1}$. Moreover, now we find a maximum tilt $\theta_{\textrm{max}}$ given by, 
\begin{equation}\label{thetamax}
\theta_{\textrm{max}}(v)=\sqrt{\frac{2gV_{sb0}f_0}{\bar{C}_{\perp} A_p}}v^{-1}, \hspace{0.5cm} {\rm{with}} \hspace{0.5cm} k S_f+A_s\ll A_p.
 \end{equation} 
The velocity $v$ and the tilting $\theta$ are to be identified with behavioural variables, unlike the rest of the parameters that are intrinsic to the fish physiology and seawater properties. However, assuming a constant swimming power, we can relate the velocity with the tilting angle.


\subsection{Constant swimming power assumption.}

The swimming power developed by the fish is $P=Tv\cos\theta$. Thus, by using Equation (\ref{Newton2}) we get $P=Dv$ and assuming a constant swimming power (CSP) we can obtain an equation that relates the horizontal velocity with the tilting,
 \beg\label{vang}
 	D(\theta,v)v=D(0,v_0)v_0=\textrm{cte},
 \en
where we have established a constant power defined by the reference value $\theta=0^\circ$ and its \textit{reference velocity} $v_0$, the maximum horizontal velocity. This expression introduces a restriction between the behavioural variables $\theta$ and $v$ that is explicitly independent of the depth $z$. It is also important to note that given that the drag coefficient depends on the velocity (see Appendix A), this equation is transcendental in $v$ and should be solved by numerical methods.

\begin{figure}
 \includegraphics[width=6cm]{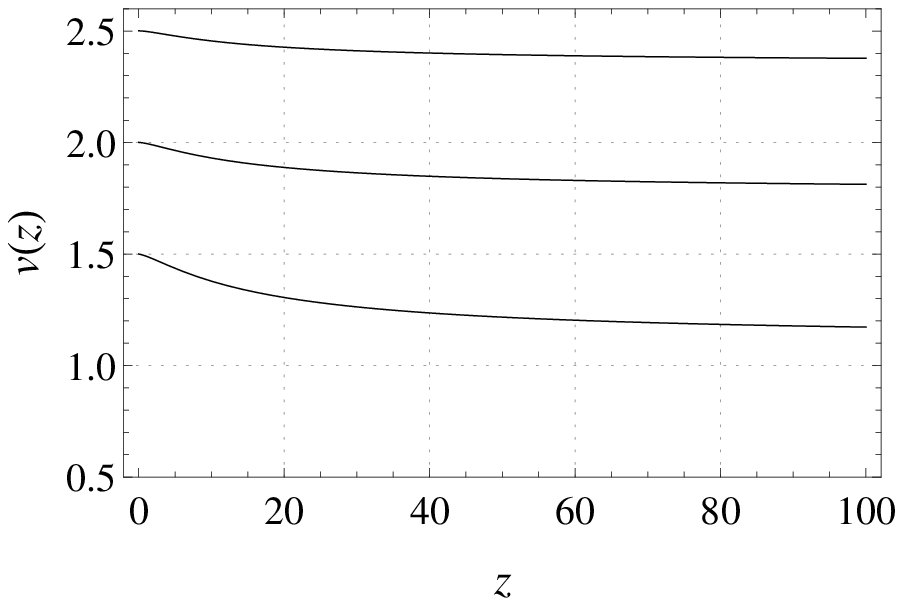}
 \includegraphics[width=6cm]{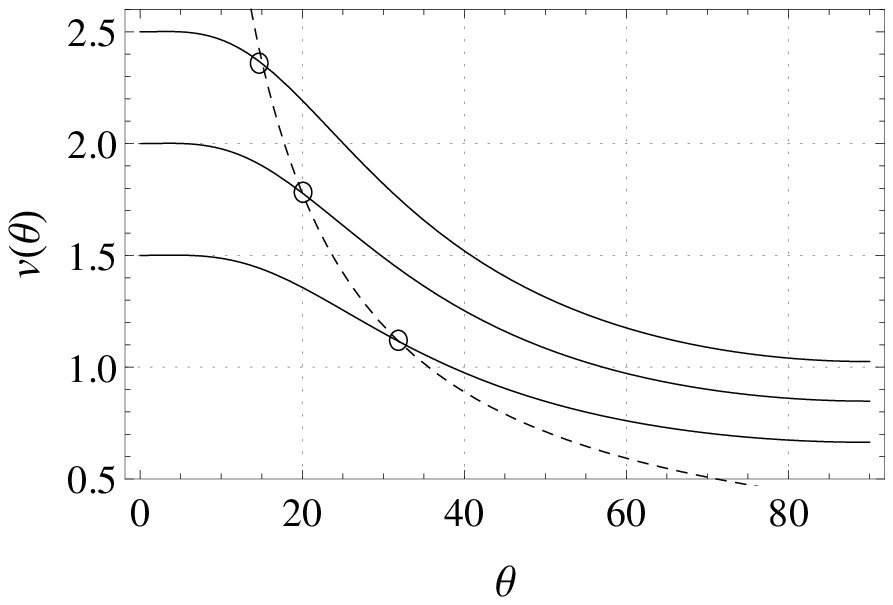}
\caption{Left plot: $v[\ell/s]\hspace{0.1cm} \rm{vs.}$ $z[m]$ for the CSP hydrodynamic model with velocities $v_0[\ell/s]=2.5, 2.0, 1.5$.
Right plot: $v[\ell/s]\hspace{0.1cm} \rm{vs.} \hspace{0.1cm} \theta[deg.]$ for the CSP hydrodynamic model with velocities $v_0[\ell/s]=2.5, 2.0, 1.5$. The dashed curve corresponds to the solution of Equation (\ref{thetamax}). The intersections marked with circles correspond to the asymptotic velocities.}
\label{Pot}
\end{figure}

As we can see in the right plot of Figure (\ref{Pot}), the maximum velocity is obtained for $\theta=0$ where all the power developed is spent on moving, whilst when swimming with tilt angle a fraction of the power is used to keep at a certain depth. 

Previously we discussed in Sec.(\ref{Tilting}) that there exists an asymptotic $\theta_{\textrm{max}}(v)$ for $z(\theta,v)$ in Equation (\ref{theta_z}). In the CSP context, this corresponds with an asymptotic decrease of the horizontal velocity with the depth up to a minimum, $v_{\text{min}}$ (see left plot of Figure (\ref{Pot})). In turn, the velocity also decays monotonically with the body tilt as in (Wilga and Lauder, 2000; Nowroozi et al., 2009; Svendsen et al., 2005; Webb, 1975). In the right plot of Figure (\ref{Pot}) we marked the intersections of the curves $v(\theta)$ with the curve $\theta_{\textrm{max}}(v)$ of Equation (\ref{thetamax}) that determine the asymptotic minimum velocities $v_{\text{min}}$ for given $v_0$ in the CSP model. Therefore, the curves to the right of the intersection points have no physical sense.

The recordings made for herrings at night by (Huse and Ona, 1996) support our results. In this case, instead of increasing swimming speed, the fish adopted a swimming tilt angle strategy. Moreover, the observed velocities were slower at deeper depths and vice versa.

We would like to remark that we are considering a stationary horizontal swimming strategy and that another swimming strategy (e.g. glide and rise) could give slower or faster velocities than this model.

\subsection{Effects of other environmental factors: the thermocline}

In this section we consider the effects on fish buoyancy and hence on fish behaviour, that could be triggered by a sharp change in the physical parameters of the water column. The particular case of a strong temperature gradient is analyzed but similar results could be expected by strong gradients on the salinity and thus affecting water density in a similar way.


\begin{figure}[htbp]
\includegraphics[width=10cm]{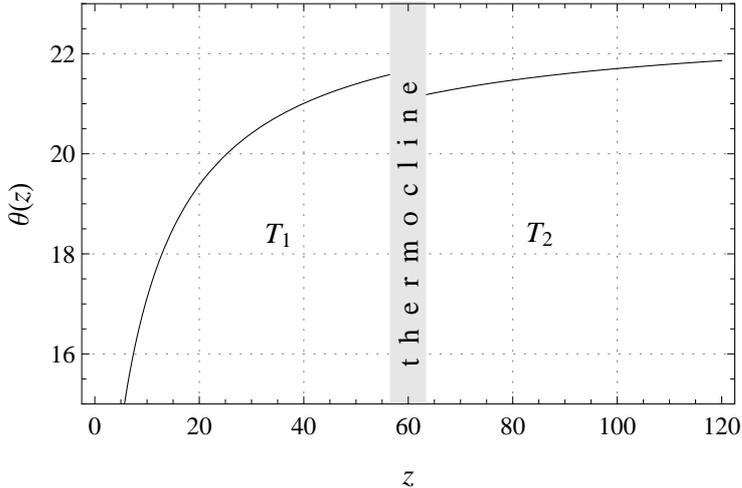}
\caption{Variation of $\theta [deg.]$ when going through the thermocline as a function of $z[m]$.}
\label{thermo}
\end{figure}

Essentially, the variables that change with temperature will be the water density $\rho(T)$ and the water viscosity $\mu(T)$ (IOC, 2010). However, in the usual thermocline temperature range the viscosity variation introduces negligible changes in the final result (see discussion of $\bar{C}_\perp$ in (\ref{BodyL})). Thus, the primarily temperature contribution will come from the change of the water density.
In Figure (\ref{thermo}) we plot the effects of a thermocline at $z\approx 60m$ with a temperature variation from $T=18^\circ C$ to $T=11^\circ C$. We observe that a small drop in the tilting ($\Delta\theta\lesssim 1^\circ C$) is produced by this temperature gradient. We can conclude that the presence of such a thermocline will not introduce significant changes in the tilting. In the same way, we presume that the variations in the salinity will not be compelling in this respect. However, the horizontal steady swimming strategy could shift to another swimming strategy when going through the thermocline. This change may occur because of behavioural reasons and not only from hydrodynamical effects. Certainly, a change in the swimming strategy may introduce bigger changes in the fish tilting than the thermocline effect on the horizontal steady swimming. For example, if the fish chooses different swimming strategies for different depths or times of the day (Huse and Ona, 1996; Paoletti and Mahadevan, 2014).

\section{An example with Argentine anchovy}\label{acoustic}

In this section we shall study the previous implications for a particular fish: Argentine anchovy. We are particularly interested in this fish because recently in Madirolas et al. (2016) there were reported in situ Target Strenght (\emph{TS}) measurements where they suggested an increasing body tilting behaviour with depth. Furthermore, in this work it was obtained a tilting vs depth behaviour by simulating theoretical \emph{TS} functions for some scenarios of swimbladder contraction and tilt angle distributions. In this sense, we wish to compare the curves $z(\theta)$ obtained from hydrodynamics considerations with the ones obtained from crossing the data of acoustic observations with simulations of \emph{TS}.

\subsection{Parameters}\label{param}

Some of the parameters used in the present example correspond to the ones in Madirolas et al. (2016) obtained from X-ray tomography for a representative specimen. On the other hand, we choose standard representative values for the seawater parameters.

\vspace{1cm}
\begin{tabular}{|c|c|c|}
\hline
Symbol & Parameter & Numerical Value \\
\hline
$\rho$ & water density & $1026 \hspace{0.1cm} kg/m^3$ \\
\hline
$\mu$ & water viscosity & $1.8\times10^{-3 } \hspace{0.1cm} kg/(m s)$ \\
\hline
$\rho_b$ & fish body density & $1060 \hspace{0.1cm} kg/m^3$ \\
\hline
$\ell$ & fish body length & $15 \hspace{0.1cm} cm$ \\
\hline
$V_{b0}$ & sea level body volume & $20 \hspace{0.1cm} cm^3$ \\
\hline
$V_{sb0}$ & sea level swimbladder volume & $0.6 \hspace{0.1cm} cm^3$ \\ \hline
$r_b$ & body transversal radius & $0.8 \hspace{0.1cm} cm$ \\
\hline
$\ka$ & pectorals lift coefficient & $0$ \\
\hline
$A_s$ & stern area & $0$ \\
\hline
$g$ & gravity acceleration & $9.81 \hspace{0.1cm} m/s^2$ \\\hline
\end{tabular}
\vspace{1cm}

We adopted the value $\kappa=0$ which implies that we are not considering lift generated by the pectoral fins ($L_f=0$). Moreover, this means the absence of any drag of the fins, $D_f=D_i=0$ (see Appendix A) . The choice of this parameter was made for simplicity reasons given that there are no measurements of $\kappa$ for this fish. Some typical values of $\kappa$ range between $\pi$ and $2\pi$, in particular for slender aerofoils like NACA 2306, 6306, etc., with $\kappa \simeq 1.4 \pi$ or NACA 6506 with $\kappa \simeq 1.2 \pi$ (Jacobs et al., 1935; Hoerner and Borst, 1975)	. If such lift fins contributions are included we found that the tilt drops about $5^{\circ}$. The difficulty of determining the value of this parameter resides primarily on the fact that it depends strongly on the fish behaviour. There is a wide range of possible fin movements: some fishes are able to control every fin ray separately, they can deform the fins or retract them, or rotate the rays at willing (Videler, 1993). Indeed, this correction appears to be meaningful, thus it would be interesting to orient some future investigations in this direction.


\subsection{Comparative with acoustic data}

The authors in  Madirolas et al. (2016) obtained an experimental curve for acoustic backscattering strength for the nighttime scattering layer of Argentine anchovy.
 \beg\label{TSexp}
    TS_{\text{exp}}(\ell,z)=31.3 \log\ell - 79.6-4.74 \log \left(1+\frac{z}{10}\right).
 \en
Besides, we used a prolate spheroid model (PSM) (Prario et al., 2015) to simulate the variation of the average TS with depth $z$ for a distribution function with mean tilt $\bar{\theta}$ and standard deviation $SD$.

The solution of the intersection of experimental curve $TS_{\text{exp}}(\ell,z)$ with the simulations $TS_{\text{sim}}(z,\bar{\theta},SD)$ (that were computed for a set of fish parameters within them $\ell=14.7 cm$, in  Madirolas et al. (2016)) yields a curve $z(\bar{\theta},SD)$ of acoustic origin.

\begin{figure}
 \includegraphics[width=6cm]{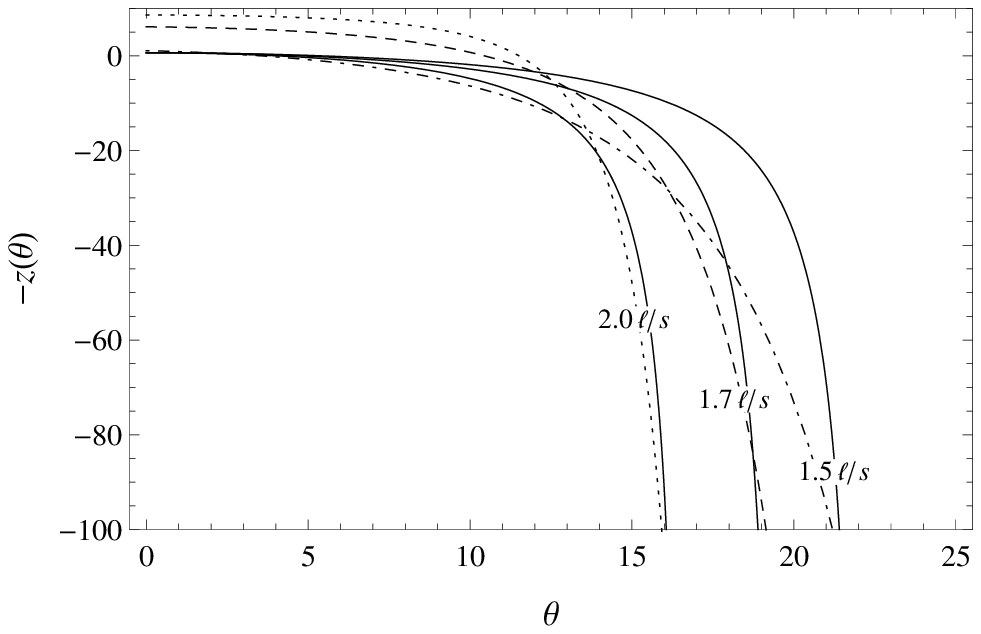}
 \includegraphics[width=6cm]{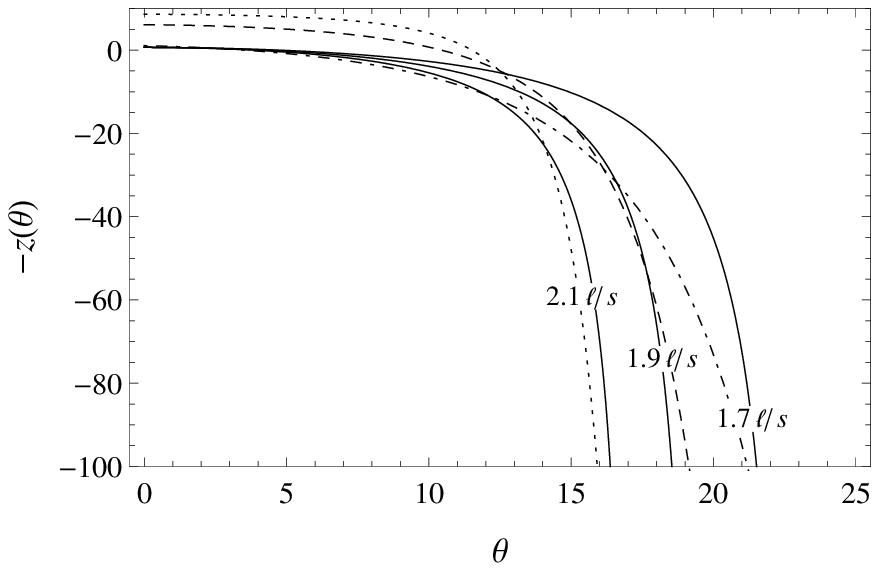}
\caption{Left plot: $-z[m]\hspace{0.1cm} \rm{vs.} \hspace{0.1cm}\theta[deg.]$ for the hydrodynamic model (black lines) with constant velocities $v[\ell/s]=2.0, 1.7, 1.5$ and from the acoustic data for $SD=10 ^\circ$ (dotted), $SD=15 ^\circ$ (dashed), $SD=20 ^\circ$ (dot-dashed).
Right plot: $-z[m]\hspace{0.1cm} \rm{vs.} \hspace{0.1cm}\theta[deg.]$ for the CSP hydrodynamic model (black lines) with velocities $v_0[\ell/s]=2.1, 1.9, 1.7$ and from the acoustic data for $SD=10 ^\circ$ (dotted), $SD=15 ^\circ$ (dashed), $SD=20 ^\circ$ (dot-dashed).}
\label{hidacu}
\end{figure}
\begin{figure}
\begin{center}
 \includegraphics[width=6
cm]{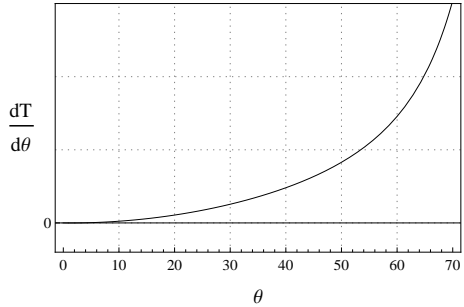}
\caption{The derivative of the thrust $dT/d\theta$ computed from Equation (\ref{Newton2}) in arbitrary units.}
\label{derivadaT}
\end{center}
\end{figure}

In the left plot of the Figure \ref{hidacu} we show $z(\theta)$ from the hydrodynamic model along the water column with constant horizontal velocities $v[\ell/s]=2.0, 1.7, 1.5$ and from the acoustic solutions for different $SD=10^\circ ,15^\circ, 20^\circ$. In the right plot of the Figure \ref{hidacu} we show $z(\theta)$ from the CSP hydrodynamic model with \textit{reference velocities} $v_0[\ell/s]=2.1, 1.9, 1.7$ and with the same acoustic data as before.
We observe that for shallower depths the hydrodynamics curves are compatible with acoustic data for $SD= 20^\circ$. However, for deeper depths the tendencies agree with a slight more polarized swimming, $SD= 10^\circ \sim 15^\circ$. 
We shall interpret this situation in terms of the change of the swimming stability with the depth. If we introduce a small perturbation $\de\theta$ from an equilibrium tilt $\theta$ (with $v$ constant)  the corresponding change in the drag $\de D$ and the lift $\de L$ shall increase for bigger $\theta$. Thus, for the fish to recover the equilibrium, it will need a stronger $\de T$ for bigger $\theta$. We show this effect in Figure (\ref{derivadaT}) where we can observe that the derivative of the thrust $T$, computed from Equation (\ref{Newton2}), it increases practically in the considered tilt range. Yet, there may exist a range of small angles where the derivative is negative, sourced by the drag term proportional to $C_\parallel \cos^3\theta$ (see Equation (\ref{CD})), which all the same appears to be residual. 
 
\section{Conclusions}\label{conclu}

We implemented a model for physostomus fish that describes the relation between the hydrodynamical forces and the body tilt during stationary swimming velocity at certain depth.

To study the swimming stability we took into account the thrust, the drag and the lift forces, and the apparent weight. The last depends on the swimbladder volume, for which we adopted Boyle's Law to describe compression with depth.
With these considerations we obtained an analytical expression for the depth as a function of the behavioural variables: body tilting and swimming velocity. We found that for rather general conditions, for fixed tilting $\theta$, the swimming depth $z$ changes hyperbolically as $v^{-2}$. Moreover, we obtained a useful approximate expression (Equation (\ref{theta_z})) for the fish tilting as a function of the depth and the swimming velocity that behaves like $v^{-1}$. Hence, for faster swimming the fish will need less body tilting for a fixed $z$. On the contrary, more dense fish will need more body tilting to cancel the extra weight. Additionally, from Equation (\ref{theta_z}) one can identify, for each swimming velocity, a maximum tilt $\theta_{\textrm{max}}$ when $z\gg z_\ast\simeq 10m$.

Then we introduced the constant swimming power hypothesis (CSP) with which we relate the velocity with the tilting without explicit depth dependence. This is a plausible biological assumption that was compared with the acoustic data. In this context, the presence of an asymptotic $\theta_{\textrm{max}}(v)$ for $z(\theta,v)$ as we previously stated, corresponds with an asymptotic decrease of the horizontal velocity with the depth. 

The temperature variation of the sea water modifies the Reynolds number, and so the body tilting of the fish may change. We analysed the changes that a thermocline would produce in the fish swimming behaviour. We discussed that the main temperature dependence comes from the density variation, while the viscosity changes with the temperature are negligible on the Reynolds number range. In spite of the presence of the temperature gradient we found that the effects of the thermocline with the body tilt are small ($\Delta\theta\approx1^\circ$ for $\Delta T\approx 7^\circ C$). Thus, we concluded that the thermocline does not affect considerably the fish tilting in the present model. 

The knowledge of the body tilt distribution for fish is of utmost importance for fisheries acoustics, given that backscattering strength is mainly originated at the exposed fish area. We tested the present hydrodynamical model against in-situ TS measurements and simulated data for Argentine Anchovy (Madirolas et al., 2016). In this case, the lack of the observed TS with depth cannot be just explained by swimbladder compression, thus we associated this difference with a body tilting excursion. From the comparison between the hydrodynamic model with the acoustic curves we found that body tilting standard deviation may show a slight decreasing trend with the depth unless a different swimming strategy is selected. We gave an interpretation of these results in terms of stability. Our results are crucial as support for more realistic modelling of TS.

Finally, we emphasize that our model is a simplification and therefore leaves out some important effects. A natural generalization of it should include, for example, a rigorous estimation of significant parameters, such as $\kappa$ in the lift fins; the incorporation of the hydrodynamic aspects of the modes swimming propulsion as well as the comparison of the results between different swimming strategies. This last point is the subject of our next work.

\section*{Acknowledgements}
We wish to thank Adrian Madirolas (INIDEP) for providing us the PSM simulated data needed to compare with our model, for useful discussions in fisheries acoustics and hospitality at INIDEP. We are also grateful with Guido Bacino (CONICET) for discussions and help in oceanography, and Constanza Brasesco (CONICET) for typos correction of the manuscript. 

\appendix

\section{Hydrodynamic forces}

\begin{figure}[htbp]
\includegraphics[width=9cm]{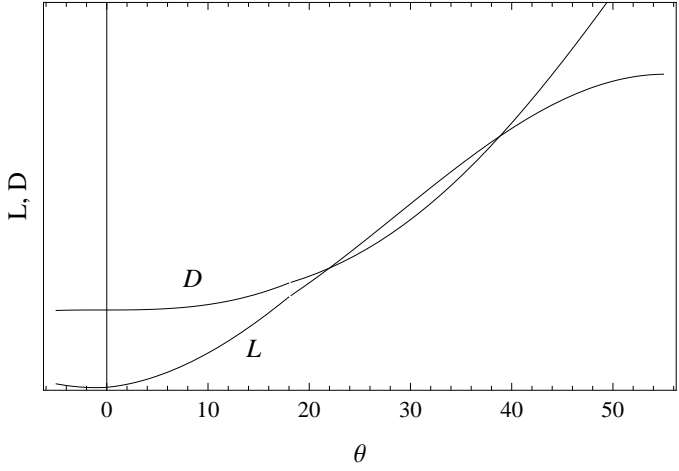}
\caption{Generic shape for $L$ and total $D$ in arbitrary units as a function of $\theta[deg.]$.}
\label{LyD}
\end{figure}

In this appendix we shall consider analytical expressions for the drag $D$ and lift $L$ forces. In Figure \ref{LyD} we plotted generic solutions for the total $D$ and $L$ that we encounter all along the paper. As we can observe, for small tilt angles the drag force dominates over the lift, but for bigger $\theta$ they both grow becoming comparable.

\subsection{Lift force}

The lift $\textbf{L}$ is produced by the dynamic effect of the water acting on the fish. It appears when an object changes the direction of flow of the water. This force is the hydrodynamic force orthogonal to the swimming direction. It helps to swim at a certain depth by stabilizing with the negative buoyant force, reaching to supply up to half of the force necessary to avoid sinking (Nowroozi et al., 2009).
We shall consider two sources of lift force, from the pectoral fins $L_f$ and from the fish body $L_b$, thus the total lift is given by
\begin{equation}\label{Lift}
L=L_{b}+L_{f}=\frac{1}{2}\rho  v^2(A_r \ C_{Lb}+ S_r\ C_{Lf}  )
\end{equation}
where $A_r$ and $S_r$ are the reference areas of the body and the fins, respectively. And $C_{Lb}$ and $C_{Lf}$ are dimensionless lift coefficients that relate the fluid variables $\rho$ and $v$ to the lift force on the body and the fins, respectively.

  \subsubsection{Body lift} \label{BodyL}
  
Jorgensen (1973) obtained some heuristic and theoretical formulas for predicting hydrodynamic forces on a streamlined body in terms of the angle of attack. The original formulas give the axial ($C_A$) and normal ($C_N$) coefficients with respect to the fish length. Yet, for our calculations we need the lift and drag coefficients, so we performed a rotation of the coordinate system at an angle $\theta$ so as to relate $C_A$ and $C_N$ with $C_L$ and $C_D$. Thus for the lift coefficient we obtain

\bega\label{CL}
    C_{Lb}&=&C_N\cos\theta-C_A\sin\theta,\\
\nonumber          &=&C_s \fr{A_{\text{s}}}{A_r}\sin{2\theta}\cos\theta\cos\fr{\theta}{2}-C_{\parallel}(v)\fr{A}{A_r}\cos^2\theta\sin\theta+ C_{\perp}(v)\fr{A_p}{A_r}\sin^2\theta\cos\theta,
 \ena
where $C_s$ is usually referred to as the apparent mass factor that depends on the ratio $L/r$, in our case $C_s\sim 1$ for streamlined bodies, $A_{\text{s}}$ is the area of the body stern, $A_r$ is the reference area,  $A_p$ is the planform area of the fish body (the maximum projected area), $A$ is the body or wetted area. Note that the global factor $A_r^{-1}$ will simplify with the factor $A_r$ in the Equation (\ref{Lift}). 

$C_{\parallel}$ is the total axial drag at zero angle of attack which is composed by friction and pressure contributions.
To estimate the value of the friction through the coefficient $C_{\text{fric}}$ we need to express it in terms of the Reynolds number (Webb, 1975), defined by
  \beg
    R_\ell=\fr{\rho \ell v}{\mu},
 \en 
where $\rho$ is the water density, $\ell$ is the fish large, $v$ the fish velocity and $\mu$ the water viscosity. Essentially there are three types of flows depending on the Reynolds number with their respective functions of the friction coefficient (Webb, 1975),
 \beg\label{Cfric}
C_{\text{fric},b}= 
\left\{%
\begin{array}{cc}
  1.33 R_\ell^{-\fr{1}{2}}, & \textsc{laminar} \\
 0.072 R_\ell^{-\fr{1}{5}}-1700 R_\ell^{-1}, & \textsc{transitional}\\
  0.072 R_\ell^{-\fr{1}{5}}, & \textsc{turbulent}\\
\end{array}%
\right.
\en
These formulas give an estimation of the friction coefficient in the laminar, transitional and turbulent flow regimes respectively. The change of the laminar to transitional regimes is governed by a Reynolds number around $5 \cdot 10^5$. In our case, $R_\ell < 10^4$, therefore we shall work in the laminar regime.

On the other hand, for streamlined bodies the pressure component can be related to the friction coefficient and to the ratio between the maximum diameter $2r$ and the length $\ell$,
 \beg
   C_{\text{press},b}=C_{\text{fric},b}G(2r/\ell),
 \en
where $G(x)=1.5x^{3/2}+7x^3$. With these expressions we can define,
 \beg
	C_\parallel=C_{\text{fric},b}+C_{\text{press},b}=C_{\text{fric},b}(1+G)
 \en

\begin{figure}
\includegraphics[width=6cm]{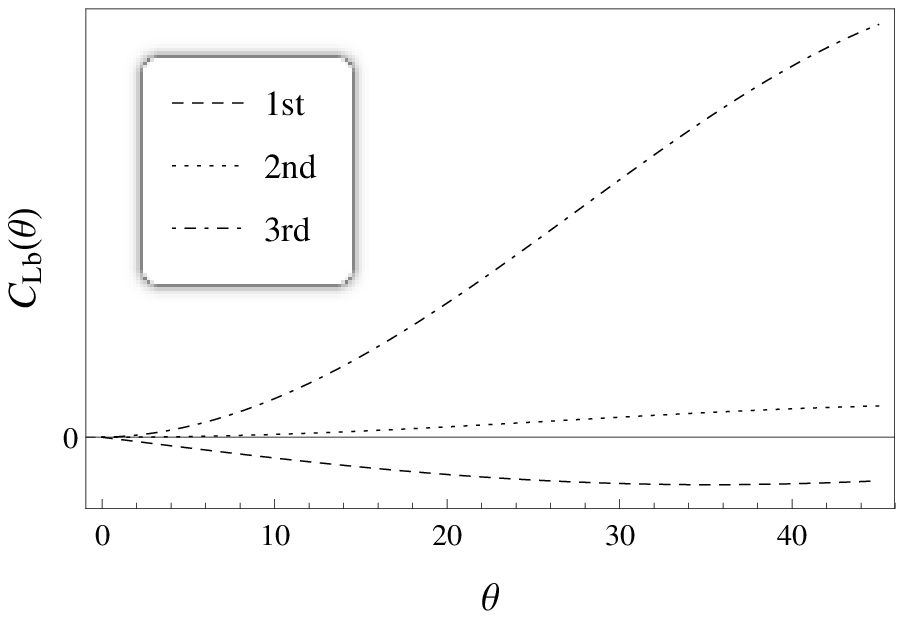}
\includegraphics[width=6cm]{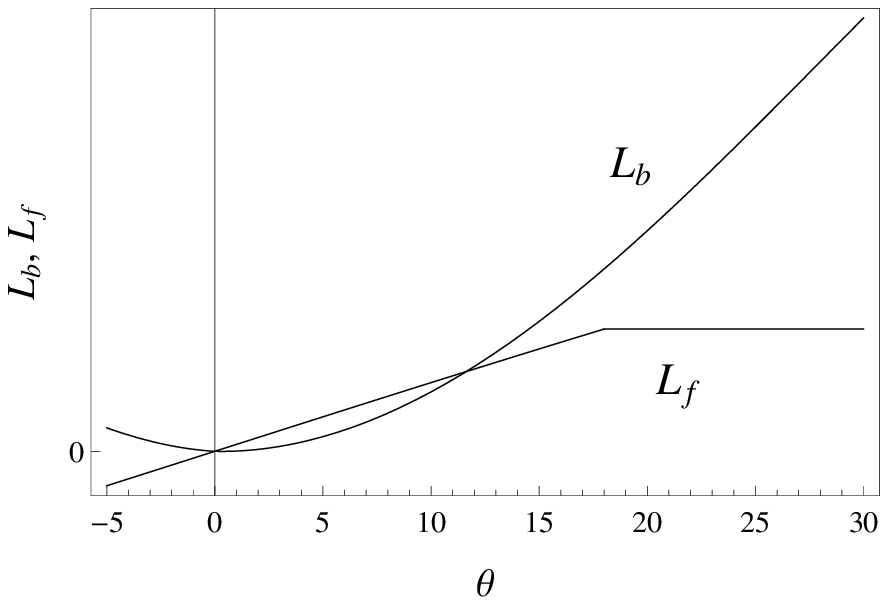}
\caption{Left plot: the different lift terms of Equation (\ref{CL}) in arbitrary units as a function of $\theta [deg.]$. Right plot: The angular dependence of body and fins lift of Equation (\ref{Lift}) in arbitrary units as a function of $\theta [deg.]$.}
\label{Lterms}
\end{figure}

$C_{\perp}$ is the cross-flow drag coefficient for a finite cylinder section, we shall use the empirical formula, $C_\perp(v)=\eta(1+10 R_N^{-2/3 })$ proposed by (White, 1991; Vogel, 1996) with a \textit{normal Reynolds number} $R_{N}=2r v \sin\theta/\mu$. We have introduced the correction factor $\eta$ for finite cylinder in the formula which in our case $\sim0.82$ (Jorgensen, 1973).
 We have tested that in the range of Reynolds number of interest, the variation of $C_\perp$ with the velocity introduces minimal changes to the final result. Thus, we keep a constant coefficient slightly bigger than $\eta$, 
  \beg
    C_\perp(v)\sim\bar{C}_\perp \gtrsim \eta.
  \en 
The first term in (\ref{CL}) corresponds to lift generated by a finite body stern area and may in general be small for streamlined fishes. The second term produces a small negative lift coming from the frontal area of the head. Lastly, in general the third term produces most of the lift that comes from the longitudinal body area exposed to cross-flow when tilting is present. These three terms are shown in the left plot of Figure \ref{Lterms}, while in the right plot we compare the angular dependence of body and fins lift.

\subsubsection{Fins lift}

We shall use for the fins, the usual simplified expression for hydrofoils (Webb, 1975),
 \beg
    C_{Lf}=\kappa \sin \alpha,
 \en
with $\alpha$ the angle of attack with respect to the flow direction and $\kappa$ a constant that depends on the form of the fins. For a hydrofoil there exists a critical angle $\alpha_{crit}$ for which above of this value the flow becomes strongly distorted and the value of $C_L$ rapidly decays, in this situation it is said that the hydrofoil stalls. In this work we shall consider that the fish can control the stall situation by changing the angle of attack of the pectoral fins. In this sense, we assume that $\alpha=\theta$ for $\theta<\alpha_{crit}\sim \pi/10$, while for $\theta>\alpha_{crit}$ then $\alpha=\alpha_{crit}$ and the coefficient remains almost constant in a way that maximizes the lift force of the fins \footnote{Indeed, a simpler assumption would be to consider a constant lift coefficient that maximizes the lift force, $C_{Lf}\lesssim\kappa \sin \alpha_{crit}$.} (see Figure \ref{Lterms}).

\subsection{Drag force}

The drag $\textbf{D}$ is the resistance force caused by the motion of a body through a fluid. The drag force always opposes to the fish velocity. We shall study the drag on the pectoral fins acting as hydrofoils and on the rest of the body. These are also given by a similar expression as for the lift force,
 \beg\label{drag}
    D=D_b+D_f=\fr{1}{2}\rho  v^2 \left( A_{r} C_{Db}+ S_r C_{Df}\right)+D_i, \en
where $C_{Db}$ and $C_{Df}$ are the body and fins drag coefficients, respectively, and $D_i$ is the induced drag that we shall analyse later in this section.


  \subsubsection{Body drag}

\begin{figure}[h]
\includegraphics[width=6cm]{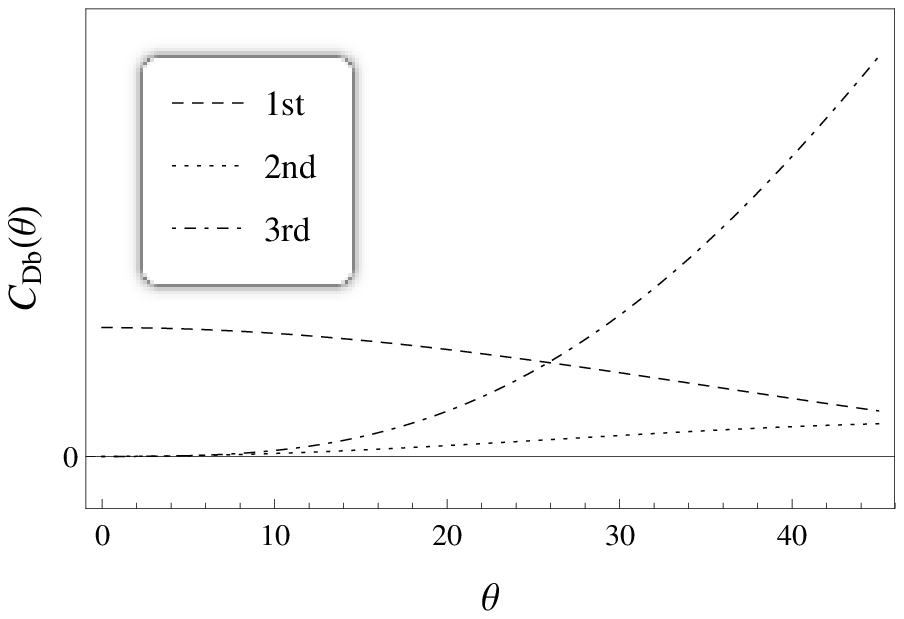}
\includegraphics[width=6cm]{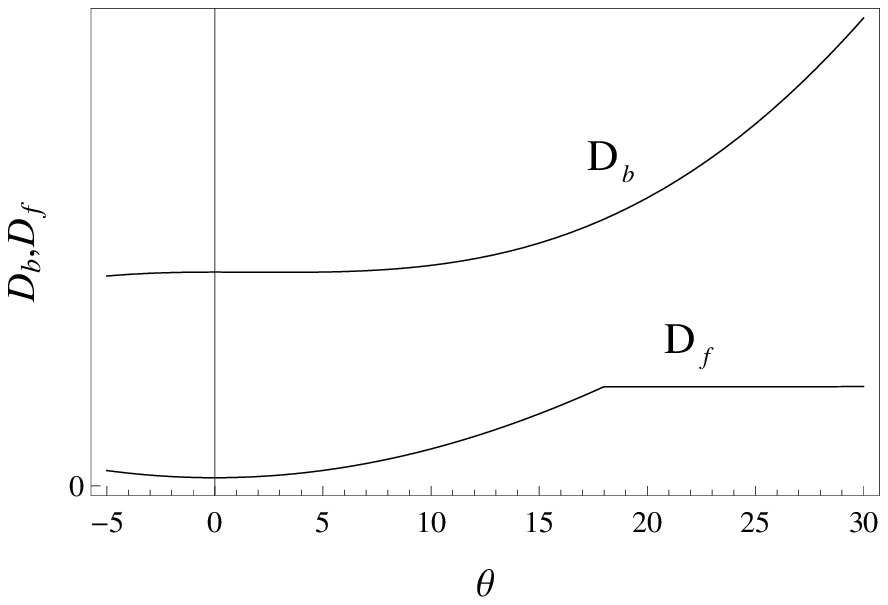}
\caption{Left plot: the different drag terms of Equation (\ref{CD}) in arbitrary units as a function of $\theta [deg.]$. Right plot: The angular dependence of body and fins drag of Equation (\ref{drag})in arbitrary units as a function of $\theta [deg.]$.}
\label{Dterms}
\end{figure}

Analogously to how the body lift coefficient was written, the body drag coefficient is related to $C_A$ and $C_N$ in the following way,
 \bega\label{CD}
    C_{Db}&=&C_N\sin\theta+C_A\cos\theta,\\
 \nonumber         &=&C_s \fr{A_s}{A_r}\sin{2\theta}\sin\theta\cos\fr{\theta}{2}+C_{\perp}\fr{A_p}{A_r}
            \sin^3\theta+C_{\parallel}\fr{A}{A_r} \cos^3\theta.
 \ena
The coefficients $C_s$, $C_\perp$ and $C_\parallel$ were defined previously in the body lift section.
These three terms are shown in the left plot of Figure \ref{Dterms}, while in the right plot we compare the angular dependence of body and fins drag.



\subsubsection{Fins drag}

The drag of the fins will be given by the sum of three terms: the friction $D_{\text{fric},f}$ and pressure drag $D_{\text{press},f}$, which speak of the direct interaction of the fluid on the fins, and the induced drag $D_{i}$, which comes whenever lift is present at hydrofoils.

The first two terms will have a friction $C_{\text{fric},f}$ and a pressure component $C_{\text{press},f}$ defined by a Reynolds number calculated with the fins chord $c$ and span $b$, thus we write,
 \beg
    D_{\text{fric},f}+D_{\text{press},f}=\fr{1}{2}\rho v^2 \left(4bc C_{\text{fric},f}+2bc \sin\theta C_{\text{press},f} \right),
 \en
where the area $4bc$ corresponds to the total wetted area of both pectoral fins, while $2bc \sin\theta$ is the area frontal to the flow direction for an angle of attack $\alpha=\theta$. Moreover, we have assumed that the contribution of the transversal area of the fins is negligible.

The magnitude of the pressure coefficient for a hydrofoil can be approximately related to the \textit{thickness ratio} $\tau$ (Webb, 1975),
 \beg
   C_{\text{press},f}=C_{\text{press},f}'+0.0056+0.01 \tau+0.1 \tau^2,
 \en
where $C_{\text{press},f}'$ is the pressure component associated to the relative camber (Webb, 1975). For our calculations we shall assume that the fins thickness ratio is very small $\tau \ll 1$, and that the camber is small also, thus $C_{\text{press},f}\simeq0.0056$. Furthermore, we obtain that the pressure drag on the pectoral fins will be much smaller than the friction drag \footnote{For a flat plate normal to the incident flow the drag force is mostly from pressure forces, and then it is independent of $R_\ell$, however this corresponds to a critical situation where the fish fins would have an angle of attack $\al\lesssim90^\circ$ which we consider a rather odd situation for steady swimming.}.

The third drag term is a consequence of the work that the hydrofoil makes on the fluid to push it downwards. Thus, $D_{i}$ is associated to the reaction of giving kinetic energy to the fluid by the fins. We have used an estimate given by momentum-jet theory (Norberg, 1990),

\beg
    D_{i}=\fr{2}{\pi b^2 \rho v^2} L_f ^2.
 \en
Using expression (\ref{drag}) we can deduce a coefficient $C_{D,i} \propto \fr{C_L^2}{AR}$, where $AR=b^2/S_{f}$ is the aspect ratio given by the relation between the fin span and the fin area.

\subsection{Demonstration of the factorization $v^2 f_1(\theta)$}

Let us consider the expression $D(\theta,v) \tan \theta+L(\theta,v)$ of the Equation (\ref{v2f1}). If we take into account the corresponding term $C_{\parallel}$ of $L_b$ and the respective of $D_b$, they cancel each other. Furthermore, as we discussed in (\ref{BodyL}) we can consider a constant $\bar{C}_{\perp}$. If we additionally consider that the viscous and pressure drag contribution from the fins is negligible and the remaining lift and drag terms are all proportional to $v^2$, then we can factorize a global factor as in Equation (\ref{v2f1}).


\end{document}